\documentclass{aa}

\usepackage{psfig}

\def\solar {\ifmmode_{\mathord\odot} \else $_{\mathord\odot}$\fi} 
\def\Msol{M\solar}
\def\simgt{\lower.5ex\hbox{$\; \buildrel\over \sim \;$}}
\def\simlt{\lower.5ex\hbox{$\; \buildrel < \over \sim \;$}}

\begin{document}

   \thesaurus{06     
              (08.02.3;  
              08.02.4;  
              08.02.6;  
              08.12.2  
              03.20.7;  
              08.12.1)  
    }

   \title{Accurate masses of very low mass stars:\\
  IV Improved mass-luminosity relations.
 \thanks{Based on observations made at the Observatoire de Haute 
   Provence (CNRS), and at the CFH Telescope, operated by the NRCC, the CNRS
   and the University of Hawaii}
}
\authorrunning{Delfosse et al.}
\titlerunning{Mass-luminosity relations of very low mass star} 
   \author{X.~Delfosse \inst{1}, T.~Forveille \inst{2,3}, 
D.~S\'egransan \inst{3}, J.-L Beuzit \inst{3,2}, S.~Udry \inst{4}, 
C.~Perrier \inst{3} and M.~Mayor \inst{4}
          }

   \offprints{Xavier Delfosse, e-mail: delfosse@ll.iac.es}

\institute{   Instituto de Astrof\'{\i}sica de Canarias
              E-38200 La Laguna, Tenerife, 
              Canary Islands, Spain
\and
              Canada-France-Hawaii Telescope Corporation, 
              65-1238 Mamaloha Highway,
              Kamuela, HI 96743, 
              U.S.A.
\and
              Observatoire de Grenoble,
              414 rue de la Piscine,
              Domaine Universitaire de S$^{\mathrm t}$ Martin d'H\`eres,
              F-38041 Grenoble,
              France
\and
              Observatoire de Gen\`eve,
              51 Ch des Maillettes,
              1290 Sauverny,
              Switzerland   
}

   \date{Received 21 July 2000; Accepted 28 September 2000}

   \maketitle

   \begin{abstract}

We present improved visual and near-infrared empirical 
mass-luminosity relations for very low mass stars (M$<$0.6~\Msol). 
These relations make use of all stellar masses in this range known
with better than 10\% accuracy, most of which are new 
determinations with 0.2 to 5\% accuracy from our own programme, presented
in a companion paper. 

As predicted by stellar structure models, the metallicity dispersion
of the field populations induces a large scatter 
around the mean V band relation, while the infrared
relations are much tighter. The agreement of the observed infrared
mass-luminosity relations with the theoretical relations of 
Baraffe et al. (\cite{baraffe98}) and Siess et al. 
(\cite{siess00}) is 
impressive, while we find an increasingly significant 
discrepancy in the V band for decreasing masses.
The theoretical mass-luminosity relation which is insufficiently steep,
and has introduced significant errors in the local stellar mass functions 
derived from V band luminosity functions.

   \end{abstract}

%

\section{Introduction}
The mass of a star is arguably its most basic characteristic, since
most stellar properties have a very steep mass dependency. Yet,
it  can only be directly determined for some stars in multiple systems,
and for most stars it has to be inferred from more directly 
observable parameters, the luminosity, the chemical composition, and the 
evolutionary status. An accurately calibrated Mass-Luminosity 
(hereafter M/L) relation, or in full generality a 
Mass-Composition-Age/Luminosity relation, is therefore
an essential astrophysical tool (e.g. Andersen \cite{andersen91}; 
\cite{andersen98}). 
It is needed at many places to convert observable stellar light to the 
underlying mass, and perhaps most importantly, to derive stellar mass 
functions from more readily obtained luminosity functions. The latter 
in turn provide an essential diagnostic of star formation theories. They 
also represent a basic building block for galactic and stellar cluster 
dynamical models.


The M/L relation is fairly well constrained for solar-type and intermediate 
mass stars: a sizeable number of such stars in eclipsing systems have had 
their mass determined with better than 1\% relative accuracy 
(Andersen \cite{andersen91}), and the theory of these stars approximately 
matches this excellent precision, when both evolutionary effects and 
metallicity are taken 
into account (Andersen \cite{andersen98}). For both smaller
and more massive stars however, the M/L relation is significantly more 
uncertain, as theory and observations meet there with new difficulties.

Here we address the low-mass end of the HR diagram, below 0.6{\Msol}, where
stellar models face two major hurdles (Chabrier \& Baraffe \cite{chabrier00},
for a recent review): 
\begin{itemize}
\item{} the onset of low temperature electron degeneracy in the stellar core
(Chabrier \& Baraffe \cite{chabrier97}; \cite{chabrier00});
\item{} a complex cold and high gravity stellar atmosphere, dominated by 
molecular and dust opacity (Allard et al. \cite{allard97}; 
Jones \& Tsuji \cite{jones97}; Allard \cite{allard98}). 
\end{itemize}
Much progress has been made in the past few years, with the realization
that an accurate atmosphere description {\it must} be used as an outer 
boundary condition to the stellar interior equations 
(Baraffe et al. \cite{baraffe95}), 
and with an increased sophistication of the atmospheric models 
(Allard et al., in prep). State of the art models 
(Baraffe et al. \cite{baraffe98}; Chabrier et 
al. \cite{chabrier00b}) now produce good to fair agreement with most 
observational 
colour-magnitude diagrams (Goldman et al. \cite{goldman99}, for an
example). 
This lends considerable credence to their general reliability. 
Their description of some of the input physics however remains 
incomplete or approximate: some molecular opacity sources are 
still described by relatively crude approximations, or by line lists
that remain incomplete (though vastly improved), the validity of the
mixing-length approximation in the convective atmosphere is questionable,
and atmospheric dust condensation and settling introduces new uncertainties 
at the lowest effective temperatures. The actual severity of these known
shortcomings of the models is unclear, making an independent check of 
their M/L prediction most desirable.

\begin{table*}
\tabcolsep 2.23mm
\begin{tabular}{|l|lll|l|rrrrrrrrr|} 
\hline
Name & $\pi$ & ${\sigma}({\pi})$ & Ref & Spectral & 
$B$ & $V$ & $R_c$ & $I_c$ & $J$ & $H$ & $K$ & $L$ & $L'$\\
     & (mas) & (mas) & & type &
    &     &       &       &     &     &     &   & \\
\hline \hline
Gl~65AB 
  &  373.7 & 2.7 & Yale95
  & M5.5V 
  & 13.87 & 12.00 & 10.37 & 8.31 & 6.24 & 5.67 & 5.33 & 5.00 & \\
Gl~234AB 
  & 243.7{\ } & 2.0{\ } & S\'eg00 & M4.5V 
  & 12.80 & 11.08 & 9.77 & 8.06 & 6.40 & 5.78 & 5.49 & 5.33 &  \\
YY~Gem   
  &  {\ \,}74.7  & 2.5  & Yale95    & M0V
  & 10.50 & 9.07 & 8.10 & 7.09  
  & 6.01 & 5.35 & 5.18 & 5.16 & \\ 
GJ~2069A 
  & {\ \,}78.05 & 5.69 & HIP & M3.5V
  &       & 11.89 & 10.68 & 9.09 & & & & & \\
Gl~473AB  
  & 227.9{\ }  & 4.6{\ } & Yale95 & M5V
  & 14.30 & 12.46 & 10.90 & 8.92 &  6.96 & 6.39 & 6.06 &  & 5.63 \\
Gl~570B   
  & 169.8{\ }  & 0.9{\ }  & For99    & M1V 
  & 9.57 & 8.09 & 7.09 & 5.97 &  4.75 & 4.14 & 3.93 & 3.77 & 3.67 \\
Gl~623AB    
  & 124.34 & 1.16 & HIP & M2.5V
  & 11.76 & 10.26 & 9.27 & 7.96 &  6.67 & 6.14 & 5.91 & & \\
CM~Dra    
  & {\ \,}69.2{\ }   & 2.5{\ } & Yale95 & M4.5V
  &  14.49 & 12.91 & 11.67 & 9.99 &  8.54 & 8.07 & 7.79 &  & \\
Gl~644AB  
  & 154.8{\ }  & 0.6{\ }  & S\'eg00 & M3V
  & 10.60 & 9.02 & 7.92 & 6.55 &  5.28 & 4.64 & 4.39 & 4.14 & \\
Gl~661AB         
  & 156.66 & 1.37 & Sod99 & M3.5V
  & 10.89 & 9.40 & 8.30 & 6.89 & 5.56 & 5.04 & 4.82 & 4.60 & \\
Gl~702AB         
  & 195.7{\ \,}  & 0.9{\ \,} & Sod99 & K0V 
  & 4.88 & 4.02 & 3.51 & 3.05 & 2.42 & 1.96 & 1.89 &  & \\
Gl~747AB         
  & 120.2{\ } & 0.2{\ } & S\'eg00 & M3V
  & 12.93 & 11.25 & 10.11 & 8.65 &  7.25 & 6.69 & 6.43 & & \\
Gl~791.2AB  
  &  112.9  &  0.3 & Ben00
  & M4.5V
  & 14.72 & 13.06 & 11.72 & 9.96 &  8.20 & 7.63 & 7.33 &  &  \\
Gl~831AB 
  & 117.5{\ } & 2.0{\ } & S\'eg00 & M4.5V
  & 13.68 & 12.01 & 10.71 & 9.02 &  7.29 & 6.70 & 6.42 &  & \\
Gl~860AB 
  &  247.5{\ } & 1.5{\ } & HIP
  & M3V+M4V 
  & 11.25 & 9.59 & 8.40 & 6.91 & 5.56 & 4.97 & 4.71 & 4.48 & \\
Gl~866ABC 
  & 293.6{\ } & 0.9{\ } & S\'eg00 & M5.5V
  & 14.29 & 12.33 & 10.66 & 8.62 &  6.50 & 5.91 & 5.57 & 5.22 & 5.01 \\
\hline
\end{tabular}
\caption{
Basic parameters for the M-dwarf systems with accurate masses.
The parallaxes mostly originate from the Hipparcos catalog (ESA \cite{esa97}),
the Yale General Catalog of trigonometric Parallaxes (Van Altena et al.
\cite{vanaltena95}), and from S\'egransan et al. 
(\cite{segransan00a}). In that paper we derived optimal combinations of 
orbital and astrometric parallaxes,
which often have a strong contribution from either an Hipparcos or a Yale
catalog astrometric parallax.
Some individual entries are additionally taken from Soderhjelm 
(\cite{soderhjelm99}), Forveille et al. (\cite{forveille99}), and
Benedict et al. (\cite{benedict00}).
All spectral types are from either Reid et al. (\cite{reid95}) or 
Hawley et al. (\cite{hawley97}) and refer to 
the integrated light of the system. The photometry is taken from the 
extensive homogenized compilation of Leggett (\cite{leggett92}), except for 
YY~Gem, Gl~702AB and GJ2069A. The photometry of GJ2069A is from Weiss 
(\cite{weiss91}).
The optical photometry of YY~Gem is from Kron et al. (\cite{kron57}) (RI),
Eggen (\cite{eggen68}) (UBV), Barnes et al. (\cite{barnes78})
(UBVRI), and the infrared photometry from Johnson (\cite{johnson65}), 
Glass (\cite{glass75}), and Veeder (\cite{veeder74}). The optical photometry
for Gl~702AB is from Bessel (\cite{bessel90}) and the infrared photometry from
Alonso et al. (\cite{alonso94}).
All photometry was converted to the Johnson-Cousins-CIT 
sytem adopted by Leggett (\cite{leggett92}) using the colour 
transformations listed 
in that paper. Multiple measurements for the same band were averaged
with equal weights. }
\label{table_basic}
\end{table*}

Detached eclipsing M-dwarf binaries are rare, with only three known to 
date. Most  mass determinations for Very Low Mass Stars (VLMS) are 
therefore instead obtained from visual and interferometric pairs, which
until recently have not yielded comparable precisions. The current state 
of the art empirical M/L relation for M dwarfs (Henry \& McCarthy \cite{henry93}; 
Henry et al. \cite{henry99}) as a result mostly rests on masses determined with 
5-20\% accuracy. The last two years have seen a dramatic evolution 
in this respect, with two groups breaking through
the former $\sim$5\% accuracy barrier. The first team used the 
1~mas per measurement astrometric  accuracy of the Fine Guidance Sensors 
(Benedict et al. \cite{benedict99}) on {\it HST} 
to determine system masses for three 
angulary resolved binaries with 2 to 10\% accuracy (Franz et al. \cite{franz98}; 
Torres et al. \cite{torres99}; Benedict et al. \cite{benedict00}). 
Shortly thereafter
our team demonstrated even better accuracies for masses of VLMS, of 
only 1-3\% (Forveille et al. \cite{forveille99}; Delfosse et al. 
\cite{delfosse99b}), by combining
very accurate radial velocities with precise angular
separations from adaptive optics. In a companion paper 
(S\'egransan et al. \cite{segransan00a}) we determine a 
dozen new or improved masses, 
with the same method and with accuracies that now range 
between 0.5 and 5\%. The same paper also presents improved masses for 
two of the three known eclipsing M-dwarf systems, with 0.2\% accuracy. 
Here we take advantage of this wealth of accurate new data, and reassess 
the VLMS M/L relation on a much firmer ground.

In Sect.~2 we briefly describe the sample of accurately determined
very low stellar masses. We then discuss the resulting empirical
M/L relation in Sect.~3, and compare it with theoretical
models in Sect.~4.

\begin{table*}
\tabcolsep 0.4mm
\begin{tabular}{|l|rl|rl|rl|rl|rl|rl|} 
\hline
Name & ${\Delta}V${\ \ } & Ref & ${\Delta}R${\ \ } 
     & Ref & ${\Delta}I${\ \ } & Ref  
     & ${\Delta}J${\ \ } & Ref & ${\Delta}H${\ \ } & Ref 
     & ${\Delta}K${\ \ } & Ref 
\\
\hline \hline
Gl~65AB    & 0.45 $\pm$ 0.08 & Hen99 
           &                 &     &            &    
           & 0.38 $\pm$ 0.03 & Hen93
           & 0.30 $\pm$ 0.02 & Hen93
           & 0.40 $\pm$ 0.07 & Hen93
\\
Gl~234AB   & 3.08 $\pm$ 0.05 & Hen99 
           &                 &     &            &    
           & 1.79 $\pm$ 0.30 & C94    
           & 1.63 $\pm$ 0.11 & C94     
           & 1.62 $\pm$ 0.02 & D00 
\\
YY~Gem     & 0.35 $\pm$ 0.20 & L78 
           & 0.34 $\pm$ 0.19 & L78 
           & 0.32 $\pm$ 0.27 & L78
           &                 &      
           &                 &      
           &                 &      
\\
GJ~2069A   & 0.79 $\pm$ 0.10 & D99a 
           &                 &     &            &    
           &                 &      &         & 
           &             &   
\\
Gl~473AB   & $-$0.01  $\pm$  0.05 & Hen99 
           &                 &     &            &    
           & 0.13 $\pm$ 0.04            &   Tor99
           & 0.20 $\pm$ 0.07            &   Tor99
           & 0.44 $\pm$ 0.09            &   Tor99
\\
Gl~570BC    & 1.66  $\pm$ 0.16 & S 
           &                 &     &            &    
           & 1.19  $\pm$ 0.05 & For99
           & 1.15  $\pm$ 0.05 & For99
           & 1.18 $\pm$ 0.03  & For99
\\
Gl~623AB     & 5.28 $\pm$ 0.10 & B96
           &                 &     &            &    
           & 3.28 $\pm$ 0.3{\ \,}  & Hen93
           & 2.65 $\pm$ 0.03 & Hen93
           & 2.87 $\pm$ 0.14 &   Hen93
\\
CM~Dra     
           & 0.14 $\pm$ 0.03 & S 
           &                 &   
           & 0.15 $\pm$ 0.01 & L77 
           &                 &      
           &                 & 
           &             &   
\\
Gl~644A-Bab &  $-$0.08 $\pm$ 0.05 & S 
           &                 &     &            &    
           &                 & 
           & $-$0.48 $\pm$  0.06  & D00 
           & $-$0.46 $\pm$  0.03 & D00  
           \\
Gl~644Ba-Bb  &  0.49 $\pm$ 0.10 & S 
           &                 &     &            &    
           &             &   
           &             &   
           &                 &      
\\
Gl~661AB  
           & 0.05 $\pm$ 0.05 & TYC 
           &                 &     &            &    
           & 0.41 $\pm$ 0.01   & Hen93
           & 0.46 $\pm$ 0.02   & Hen93
           & 0.42 $\pm$ 0.07   & Hen93   \\  
Gl~702AB  
           &  1.86 $\pm$ 0.02 &  TYC
           &                 &     &            &    
           & 1.51 $\pm$ 0.04 & Hen93
           &             &   
           & 0.74 $\pm$ 0.03 & Hen93 \\    
Gl~747AB   & 0.22 $\pm$ 0.05 & S 
           &                 &     &            &    
           &                 &     &            &    
           & 0.10 $\pm$ 0.03 & D00 
\\
Gl~791.2AB & 3.27 $\pm$ 0.10 & Ben00
           &                 &     &            &    
           &                 &     &            &    
           &                 &      
\\
Gl~831AB   
           & 2.10 $\pm$ 0.06 & Hen99
           &                 &     &            &    
           &                 &     
           & 1.26 $\pm$ 0.03  & D00 
           & 1.28 $\pm$ 0.02  & D00 
\\
Gl~860AB
%
   & 1.70 $\pm$ 0.09 & HIP 
           &                 &     
           &                 &    
           & 1.19 $\pm$ 0.10 & Hen93
           & 1.14 $\pm$ 0.05 & Hen93
           & 1.37 $\pm$ 0.08 & Hen93
\\
Gl~866AC-B 
           & 0.40 $\pm$ 0.10 & Hen99  %
           &                 &     &            &    
           &            &    
           & 0.52 $\pm$ 0.03 & D00 
           & 0.54 $\pm$ 0.03 & D00 
\\
Gl~866AC  & 2.04 $\pm$ 0.40 & S
           &                 &     &            &    
           &                 &     &            &    
           &                 &      
\\
\hline
\end{tabular}
\caption{Magnitude differences for M-dwarf systems with accurate masses.
Reference codes are: 
TYC for the Tycho catalogue
HIP for the Hipparcos catalogue
L77 for Lacy (\cite{lacy77}), 
L78 for Leung \& Schneider (\cite{leung78}), 
Hen93 for Henry \& McCarthy (\cite{henry93}),
C94 for Coppenbarger et al. (\cite{coppenbarger94}), 
B96 for Barbieri et al. (\cite{barbieri96}), 
Ben00 for Benedict et al. (\cite{benedict00}),
Tor99 for Torres et al. (\cite{torres99}) 
Hen99 for Henry et al. (\cite{henry99}), 
D99a for Delfosse et al. (\cite{delfosse99a}), 
For99 for Forveille et al. (\cite{forveille99}), 
and D00 for the present paper. 
S stands for spectroscopic magnitude differences, infered
from the relative line depths for double-lined spectroscopic binaries.}
\label{table_dmag}
\end{table*}

\section{Sample}

\subsection{Accurate masses for M dwarfs}

We adopt a 10\% mass accuracy cutoff for inclusion in our
new M dwarf M/L relations, as a compromise 
between good statistics and the quality of the individual
measurements. To our knowledge 32 M dwarfs fulfill this criterion.
They can be divided into four broad categories:

1. Four systems have orbits of a quality that hasn't changed much 
since Henry \& McCarthy (\cite{henry93}), but masses which are now
somewhat better determined thanks to the availability of
the Hipparcos parallaxes (Henry et al. \cite{henry99}, and this
paper): Gl~65, Gl~661, Gl~702, and Gl~860. One should note however
that the Gl~661 masses derived by Martin et al. (\cite{martin98}) 
represent a $\sim$3~$\sigma$ correction to the Henry \& McCarthy 
(\cite{henry93}) values. The new masses are much more consistent
with the average M/L relation, and we believe that Henry \& McCarthy 
(\cite{henry93}) had underestimated their standard errors for that 
particular pair.
These four systems have long to very long periods, up to 90~years for 
Gl~702AB. 
Their inclusion is a testimony to the care and dedication of 
binary star observers over many decades, but the accuracy of most of
these masses is unlikely to be significantly improved over the 
next few years.

2. The three eclipsing systems have extremely accurate masses, with 
relative precisions of 0.5\% for CM~Dra (Metcalfe et al. \cite{metcalfe96}) 
and 0.2\% 
for YY~Gem and GJ~2069A (S\'egransan et al. \cite{segransan00a}). 
Their parallaxes are 
unfortunately less precisely known, in part due to their slightly 
larger distances than those of the visual systems. Their 
luminosities are therefore more uncertain than their masses. Their
flux ratios in the near-IR JHK bands have also not yet been 
determined.

3. The masses of Gl~473 (8\%; Torres et al. \cite{torres99}) and Gl~791.2 
(2\%; Benedict et al. \cite{benedict00}) result from the effort 
of the FGS astrometry 
team on HST (Benedict et al. \cite{benedict99}). 
We have chosen to temporarily exclude 
the measurement of the Gl~748 system mass by the same group (5\%; 
Franz et al. \cite{franz98}): a M/L relation has to be used to estimate 
individual masses for the two stars in that system. This would introduce 
an undesirable 
circular aspect to our discussion. We have on the other hand 
retained their Gl~473 determination, which formally has the same 
weakness. The two components of that system are sufficiently similar 
that this contributes negligible additional uncertainties.

4. Most of the masses in Table~\ref{mass} result from our 
own programme: since 1995 we have been monitoring a sample of solar 
neighbourhood M~dwarfs with high precision radial velocity and
adaptive optics imaging observations (Delfosse et al. \cite{delfosse99c}, 
for a complete
presentation of the project). These observations have resulted in a new
orbit for Gl~747, and in improved orbits and masses with 0.5 to 5\% accuracy 
for Gl~234, Gl~644, Gl~831, Gl~866 (S\'egransan et
al., \cite{segransan00a}), Gl~570B and Gl~623 (S\'egransan et al., 
in prep). Some additionnal binaries, including discoveries 
from that programme (Delfosse et al. 
\cite{delfosse99c}, Beuzit et al., in prep.) are nearing the time 
when their masses will be known
with similar accuracies.

\begin{figure*}
\begin{tabular}{cc}
\tabcolsep 0.2mm
\psfig{height=6.2cm,file=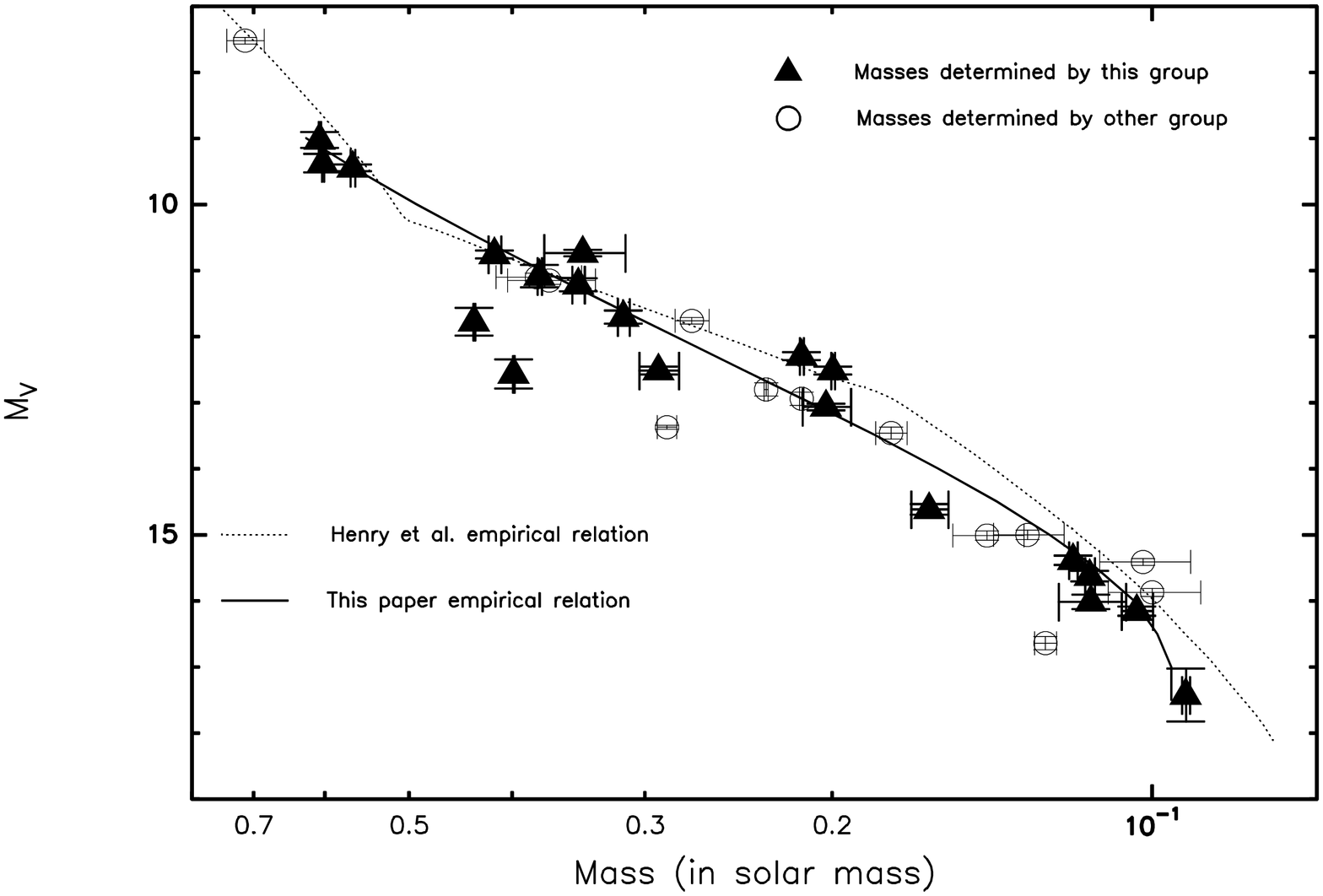,angle=-90} &
\psfig{height=6.2cm,file=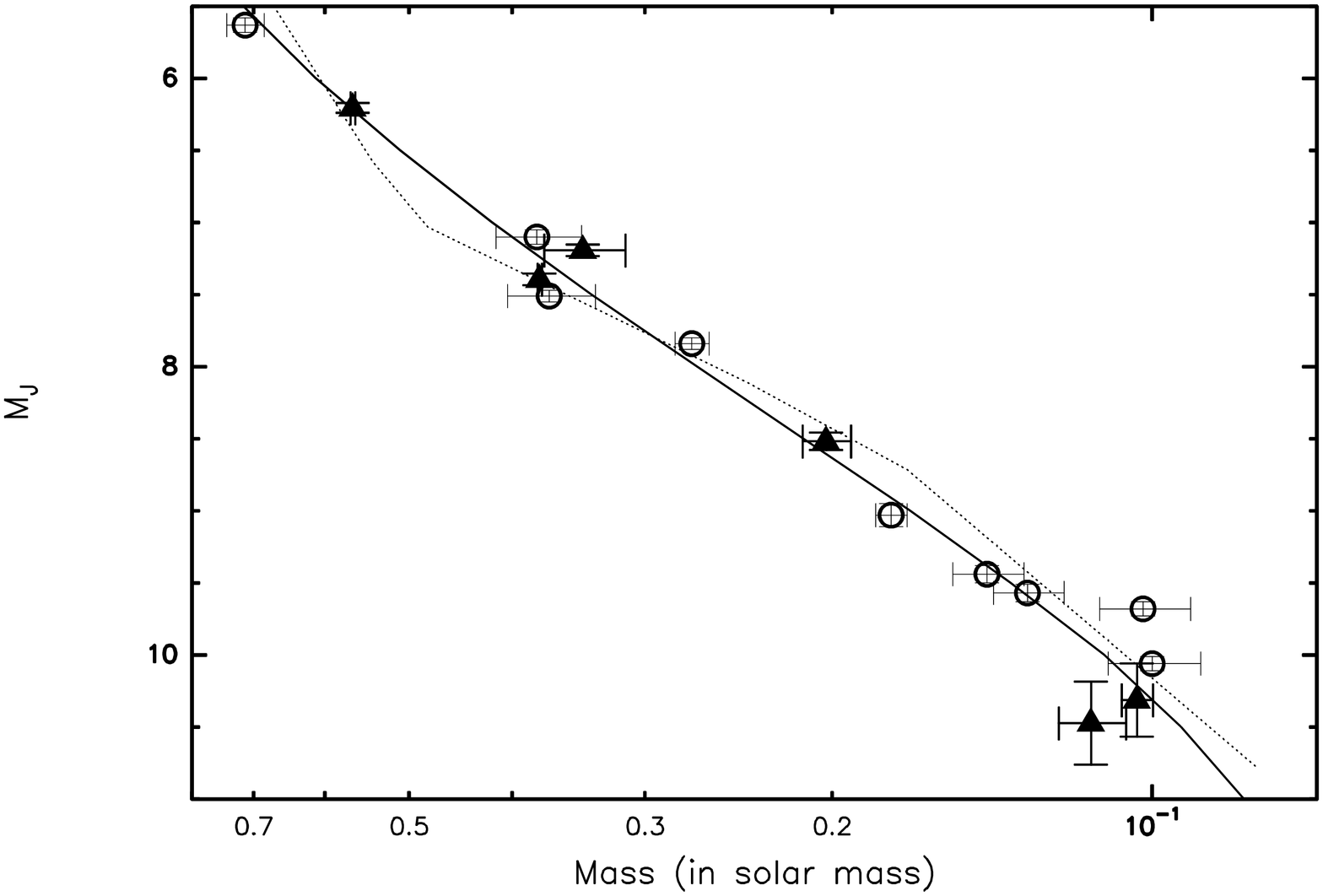,angle=-90} \\
\psfig{height=6.2cm,file=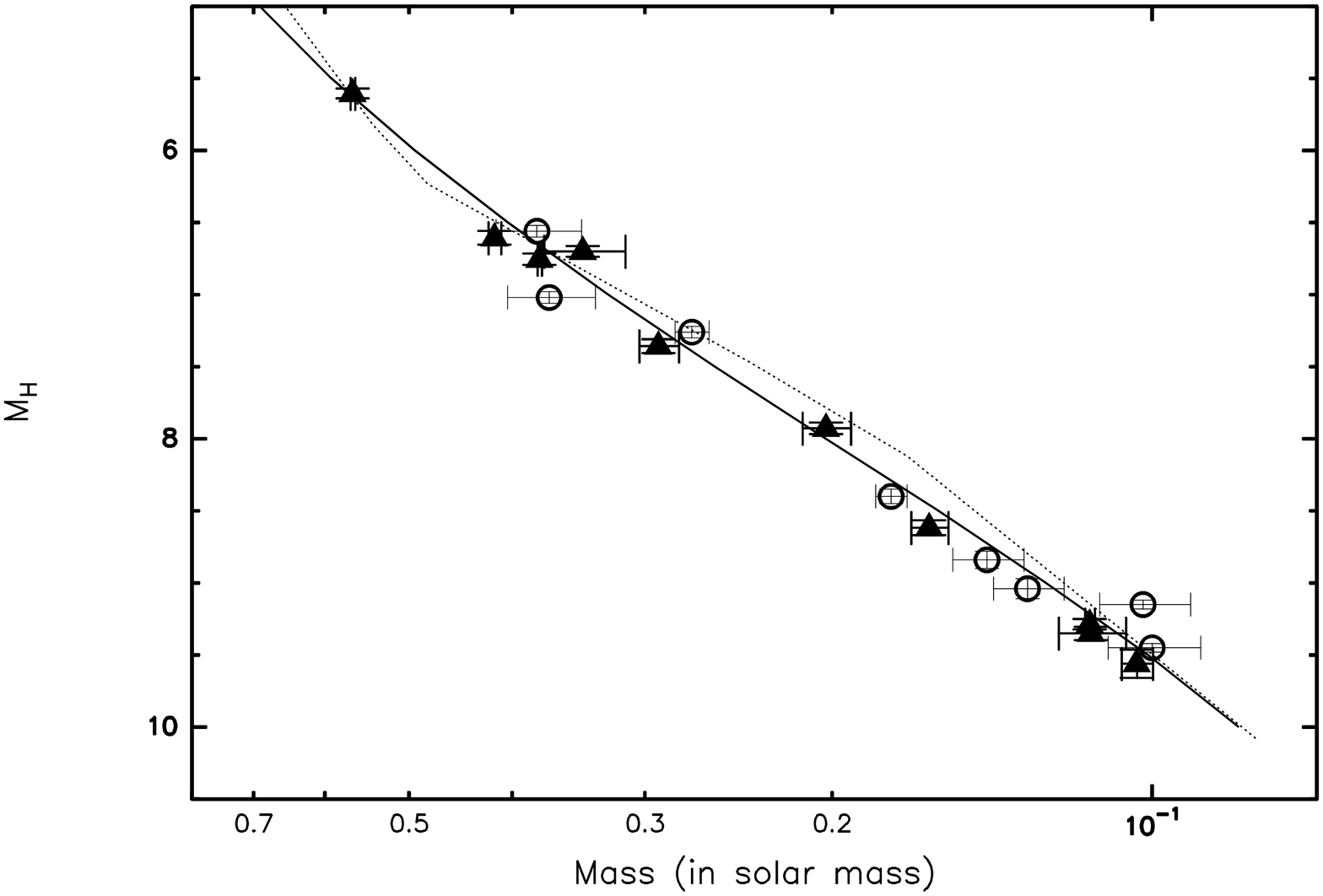,angle=-90} &
\psfig{height=6.2cm,file=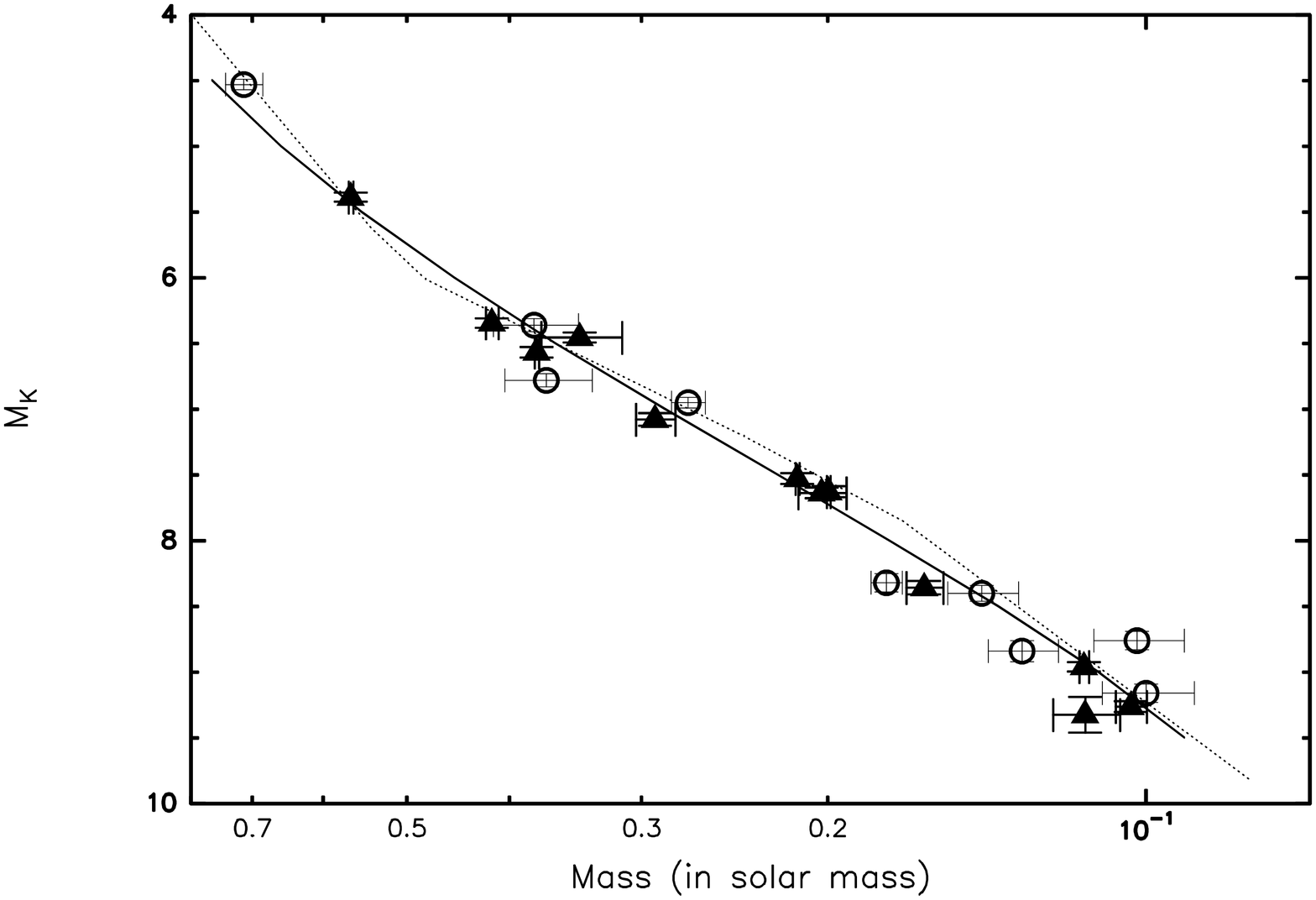,angle=-90} \\
\end{tabular}
\caption{V, J, H and K band M/L relations. 
The circles are data from Henry \& McCarthy (\cite{henry93}), 
Torres et al. (\cite{torres99}), 
Henry et al. (\cite{henry99}), Benedict et al. (\cite{benedict00}) and
Metcalfe et al. (\cite{metcalfe96}). The triangles 
represent our recent mesurements (S\'egransan et al. \cite{segransan00a}; 
S\'egransan et al., in prep.; 
and Forveille et al. \cite{forveille99}). 
The masses and 
luminosities used in this figure are also listed in Table~\ref{mass}.
The two curves represent the piecewise linear relation of Henry \& 
McCarthy (\cite{henry93}; dotted line) and our polynomial fit (solid 
line).}
\label{mass-lum}
\end{figure*}

\begin{table*}
\tabcolsep 2.7mm
\begin{tabular}{|ll|ll|llll|} \hline
\multicolumn{2}{|l|}{Name} & Mass (in \Msol) & Ref & M$_{\rm V}$ & M$_{\rm J}$ & M$_{\rm H}$ &M $_{\rm K}$  \\ \hline
Gl 65  & A &  0.102{\ \,} $\pm$ 0.010{\ \,}  (9.8\%) & Hen99 & 15.41 $\pm$ 0.05 
  &  {\ \,}9.68 $\pm$ 0.05 & 9.15 $\pm$ 0.03 & 8.76 $\pm$ 0.07 \\  
       & B &  0.100{\ \,} $\pm$ 0.010{\ \,} (10.0\%) & Hen99 & 15.87 $\pm$ 0.06 
  & 10.06 $\pm$ 0.05 & 9.45 $\pm$ 0.03 & 9.16 $\pm$ 0.07 \\
Gl 234 & A & 0.2027 $\pm$ 0.0106 (5.2\%) & Seg00a &13.07 $\pm$ 0.05 
  &  8.52 $\pm$ 0.06 & 7.93 $\pm$ 0.04 & 7.64 $\pm$ 0.04 \\
       & B & 0.1034 $\pm$ 0.0035 (3.4\%) & Seg00a &16.16 $\pm$ 0.07 
  & 10.31 $\pm$ 0.25 & 9.56 $\pm$ 0.10 & 9.26 $\pm$ 0.04 \\
YY Gem & a & 0.6028 $\pm$ 0.0014 (0.2\%) & Seg00a & {\ \,}9.03 $\pm$ 0.12 
  & & &  \\
       & b & 0.6069 $\pm$ 0.0014 (0.2\%) & Seg00a & {\ \,}9.38 $\pm$ 0.14 
  & & &  \\
GJ 2069A& a & 0.4344 $\pm$ 0.0008 (0.2\%) & Seg00a & 11.78 $\pm$ 0.18 & & & \\
       & b & 0.3987 $\pm$ 0.0007 (0.2\%) & Seg00a & 12.57 $\pm$ 0.19 & & & \\
Gl 473 & A & 0.143{\ \,}  $\pm$ 0.011{\ \,}  (7.7\%) & Tor99 
  & 15.01 $\pm$ 0.07 
  & {\ \,}9.44 $\pm$0.06 & 8.84 $\pm$ 0.06 & 8.40 $\pm$ 0.06 \\
       & B & 0.131{\ \,}  $\pm$ 0.010{\ \,}  (7.6\%) & Tor99 
  & 15.00 $\pm$ 0.07 
  & {\ \,}9.57 $\pm$ 0.06 & 9.04 $\pm$ 0.07 & 8.84 $\pm$ 0.08 \\
Gl 570 & B & 0.5656 $\pm$ 0.0029 (0.5\%) & For99 & {\ \,}9.45 $\pm$ 0.05 
  & {\ \,}6.21 $\pm$ 0.03 & 5.61 $\pm$ 0.03 & 5.39 $\pm$ 0.03 \\
       & C & 0.3770 $\pm$ 0.0018 (0.5\%) & For99 & 11.09 $\pm$ 0.17 
  & {\ \,}7.40 $\pm$ 0.04 & 6.76 $\pm$ 0.04 & 6.57 $\pm$ 0.04 \\
Gl~623 & A & 0.3432 $\pm$ 0.0301 (8.8\%) & Seg00b & 10.74 $\pm$ 0.05 
  & {\ \,}7.19 $\pm$ 0.04 & 6.70 $\pm$ 0.04 & 6.46 $\pm$ 0.04 \\
       & B & 0.1142 $\pm$ 0.0083 (7.3\%) & Seg00b & 16.02 $\pm$ 0.11 
  & 10.47 $\pm$ 0.29 & 9.35 $\pm$ 0.05 & 9.33 $\pm$ 0.14 \\
CM Dra & a & 0.2307 $\pm$ 0.0010  (0.5\%) & Met96 & 12.80 $\pm$ 0.1{\ \,} 
  & & & \\
       & b & 0.2136 $\pm$ 0.0010  (0.5\%) & Met96 & 12.94 $\pm$ 0.1{\ \,} 
  & & & \\
Gl 644 & A & 0.4155 $\pm$ 0.0057 (1.4\%) & Seg00a & 10.76 $\pm$ 0.06 
  & & {\ \,}6.61 $\pm$ 0.05 & 6.35 $\pm$ 0.04 \\
       & Ba & 0.3466 $\pm$ 0.0047 (1.3\%) & Seg00a & 11.22 $\pm$ 0.10   & & & \\
       & Bb & 0.3143 $\pm$ 0.0040 (1.3\%) & Seg00a & 11.71 $\pm$ 0.10  & & & \\
Gl 661 & A & 0.379{\ \,}  $\pm$ 0.035{\ \,}  (9.2\%) & Mar98 
  & 11.10 $\pm$ 0.06 
  & {\ \,}7.10 $\pm$ 0.05 & 6.56 $\pm$  0.04 & 6.36 $\pm$ 0.05 \\
       & B & 0.369{\ \,}  $\pm$ 0.035{\ \,}  (9.5\%) & Mar98 
  & 11.15 $\pm$ 0.06 
  & {\ \,}7.51 $\pm$ 0.04 & 7.02 $\pm$  0.04 & 6.78 $\pm$ 0.05 \\
Gl 702 & B & 0.713{\ \,}  $\pm$ 0.029{\ \,}  (4.1\%) & Hen93  
  & {\ \,}7.52 $\pm$ 0.05
  & {\ \,}5.63 $\pm$ 0.05 & & 4.53 $\pm$ 0.04  \\
Gl 747 & A & 0.2137 $\pm$ 0.0009 (0.4\%) & Seg00a & 12.30 $\pm$ 0.06 
  & & & 7.53 $\pm$ 0.04 \\
       & B & 0.1997 $\pm$ 0.0008 (0.4\%) & Seg00a & 12.52 $\pm$ 0.06 
  & & & 7.63 $\pm$ 0.04 \\
Gl 791.2 & A & 0.286{\ \,} $\pm$ 0.006{\ \,} (2.1\%) & Ben00 & 13.37 $\pm$ 0.03 
  & & & \\
       & B & 0.126{\ \,} $\pm$ 0.003{\ \,} (2.4\%) & Ben00 & 16.64 $\pm$ 0.10 & & &  \\
Gl 831 & A & 0.2913 $\pm$ 0.0125 (4.3\%) & Seg00a & 12.52 $\pm$ 0.06
  & & 7.36 $\pm$ 0.05 & 7.08 $\pm$ 0.05 \\
       & B & 0.1621 $\pm$ 0.0065 (4.0\%) & Seg00a & 14.62 $\pm$ 0.08 
  & & 8.62 $\pm$ 0.05 & 8.36 $\pm$ 0.05 \\
Gl 860 & A & 0.2711  $\pm$ 0.0100  (4.3\%) & Hen93 & 11.76 $\pm$ 0.05 
  & {\ \,}7.84 $\pm$ 0.04 & 7.26 $\pm$ 0.04 & 6.95 $\pm$ 0.04 \\
       & B & 0.1762  $\pm$ 0.0066  (4.7\%) & Hen99 & 13.46 $\pm$ 0.09 
  & {\ \,}9.03 $\pm$ 0.08 & 8.40 $\pm$ 0.05 & 8.32 $\pm$ 0.07 \\
Gl 866 & A & 0.1187 $\pm$ 0.0011 (0.9\%) & Seg00a & 15.39 $\pm$ 0.07  & & & \\
       & B & 0.1145 $\pm$ 0.0012 (1.0\%) & Seg00a & 15.64 $\pm$ 0.08 
  & & 9.29 $\pm$ 0.04 & 8.96 $\pm$ 0.04 \\
       & C & 0.0930 $\pm$ 0.0008 (0.9\%) & Seg00a & 17.43 $\pm$ 0.40  & & & \\ 
\hline
\end{tabular} 
\caption{Masses and absolute magnitudes for the M dwarfs used in the
M/L relation. The mass references are 
Met96 for Metcalfe et al. (\cite{metcalfe96}), 
Hen99 for Henry et al. (\cite{henry99}),
Hen93 for Henry \& McCarthy (\cite{henry93}), 
Mar98 for Martin et al. (\cite{martin98}),
Tor99 for Torres et al. (\cite{torres99}), 
Ben00 for Benedict et al. (\cite{benedict00}),
Seg00a for S\'egransan et al. (\cite{segransan00a}), 
Seg00b for S\'egransan et al. (in prep.) and 
For99 for Forveille et al. (\cite{forveille99}). When relevant we have 
modified the masses from Henry \& McCarthy (\cite{henry93}) and Henry 
et al. (\cite{henry99}) to reflect a more accurate parallax 
in Table~\ref{table_basic} than was available to these authors. The individual
absolute magnitudes are determined from the system magnitudes and parallaxes
listed in Table~\ref{table_basic}, with the magnitude differences of 
Table~\ref{table_dmag}.
} 
\label{mass}
\end{table*}

\subsection{Photometry}

Table \ref{table_basic} lists the basic properties of the selected
systems: parallaxes, spectral types and integrated photometry. 
Table~\ref{table_dmag} lists the magnitude differences for the V, R, I,
J, H and K photometric bands. The near-IR flux ratios were obtained
either from litterature infrared speckle observations, or extracted from 
our adaptics images (S\'egransan et al. \cite{segransan00a}). For the three 
eclipsing systems the visible flux ratios were adopted from analyses of
the light curves (Leung \& Schneider \cite{leung78}; Delfosse et al. 
\cite{delfosse99a}; Lacy \cite{lacy77}).
For the visual binaries, they were preferentially adopted from the 
FGS work of Henry et al. (\cite{henry99}), with the standard errors quoted in that
article. When such measurements were unavailable, we relied instead on 
spectroscopic magnitude differences, using the relative areas of the
ELODIE cross-correlation peaks as a proxy for the V band magnitude
difference. The ELODIE cross-correlation has an effective bandpass
centered close to the central wavelenghth of the Johnson V filter, 
but it is significantly broader. A "colour-tranformation"
would thus in principle be needed to derive V band magnitude differences.
A comparison with the direct measurements of 
Henry et al. (\cite{henry99}) for the 
sources in common shows maximum relative errors of $\sim$10\% from
neglecting this transformation: a 0.5~magnitude contrast is in error 
by at most 0.05 magnitude, and a 
2~magnitudes one by at most 0.2 magnitude. We have therefore adopted
the larger of 0.05~magnitude and 10\% of the 
magnitude difference as a conservative estimate of the standard error 
for these spectroscopic magnitude differences.


\section{Visible and infrared Mass/Luminosity relations}

The masses are listed in Table~\ref{mass}, with the individual absolute 
magnitudes derived from Table~\ref{table_basic} and Table~\ref{table_dmag}
for the four photometric bands (V, J, H and K) which have significant 
numbers of measurements. 
Fig.~\ref{mass-lum} shows the M/L relations for these 4 photometric bands.
As can be seen immediately in Fig.~\ref{mass-lum},  $\sim$20~stars 
define the V and K relations, while the J and H ones still have smaller 
numbers of stars. A number of systems still lack magnitude difference 
measurements in those two bands. 

\begin{figure}
\psfig{height=6.2cm,file=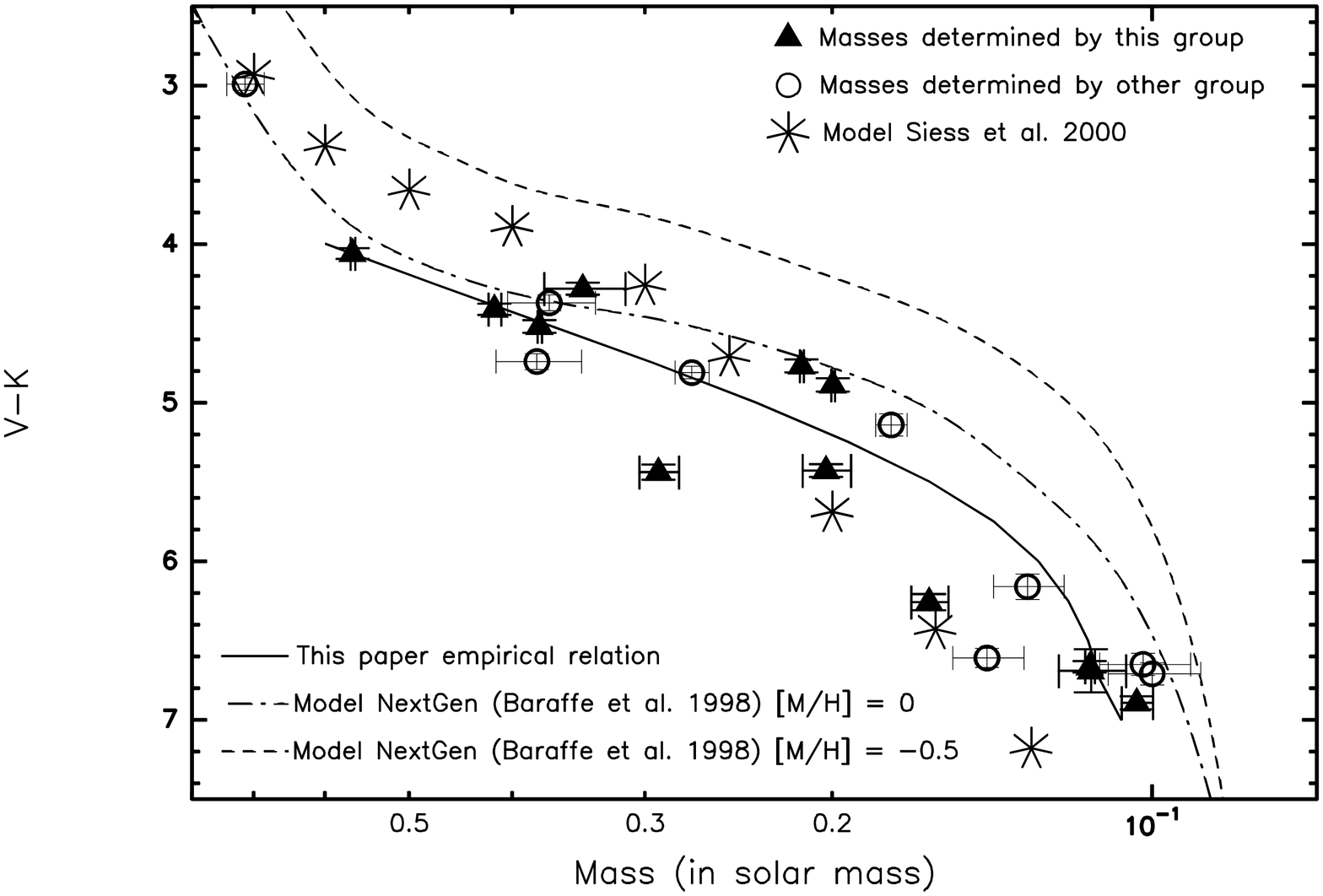,angle=-90} 
\caption{mass-colour (V-K) relation for M dwarfs.
The three curves are 5~Gyr theoretical isochrones from 
Baraffe et al. (\cite{baraffe98}) 
for two metallicities and our polynomial fit. The Siess et al. 
(\cite{siess00}) model are represented for 5~Gyr and solar
metallicity with asterisks.}
\label{mass-spe}
\end{figure}

Figure~\ref{mass-spe} presents the relation between stellar mass
and the V-K colour index. This relation probably has 
too large an intrinsic dispersions to be generally useful, and is provided 
here mostly for illustration, and as a warning to potential users of similar 
relations.

Figure~\ref{mass-lum} shows the piecewise-linear relations adjusted by 
Henry \& McCarthy (\cite{henry93}) to the J, H and K band data then
available to them, and their piecewise-quadratic relation for the V
band, with its Henry et al. (\cite{henry99}; V band) update for the
lower masses. These relations provide a reasonable description of 
the new data, but they do show significant discrepancies, in particular
around their breakpoints. Clearly the quality of the new masses 
warrants the use of higher order polynomials. We have found that the
following fourth degree polynomials provide good descriptions of the 
data in Figs.~\ref{mass-lum} and \ref{mass-spe}: 

\begin{displaymath}
\begin{array}{l}
log(M/{\Msol})=10^{-3}{\times}[0.3+1.87{\times}{M_V}
+7.6140{\times}{M_V}^2\\
-1.6980{\times}{M_V}^3
+0.060958{\times}{M_V}^4] {\ \ \ }{\mathrm{for}} {\ }{M_V} 
\in [9,17]
\end{array}
\end{displaymath}

\begin{displaymath}
\begin{array}{l}
log(M/{\Msol})=10^{-3}{\times}[1.6+6.01{\times}{M_J}
+14.888{\times}{M_J}^2 \\
-5.3557{\times}{M_J}^3
+2.8518.10^{-4}{\times}{M_J}^4] {\ \ \ }{\mathrm{for}} {\ }{M_J}
\in [5.5,11]
\end{array}
\end{displaymath}

\begin{displaymath}
\begin{array}{l}
log(M/{\Msol})=10^{-3}{\times}[1.4+4.76{\times}{M_H}
+10.641{\times}{M_H}^2 \\
 -5.0320{\times}{M_H}^3 +0.28396{\times}{M_H}^4]
{\ \ \ }{\mathrm{for}} {\ }{M_H}
\in [5,10]
\end{array}
\end{displaymath}

\begin{displaymath}
\begin{array}{l}
log(M/{\Msol})=10^{-3}{\times}[1.8+6.12{\times}{ M_K}
+13.205{\times}{ M_K}^2\\
 -6.2315{\times}{ M_K}^3+0.37529{\times}{ M_K}^4
{\ \ \ }{\mathrm{for}} {\ }{ M_K}
\in [4.5,9.5]
\end{array}
\end{displaymath}

\begin{displaymath}
\begin{array}{l}
log(M/{\Msol})=10^{-3}{\times}[7.4+17.61{\times}{ (V-K)}\\
+33.216{\times}{ (V-K)}^2
+34.222{\times}{ (V-K)}^3 \\
-27.1986{\times}{ (V-K)}^4
+4.94647{\times}{ (V-K)}^5\\
-0.27454{\times}{ (V-K)}^6
{\ \ \ }{\mathrm{for}} {\ }{ V-K} \in [4,7]
\end{array}
\end{displaymath}


One striking characteristic of Fig.~\ref{mass-lum} is the
very different scatters in its V and K diagrams. 
The V plot displays considerable dispersion around a mean relation, and some
of its best measurements are also some of the most discrepant. 
The K plot on the other hand shows a one
to one relation between Mass and Luminosity, and its (mild) outliers 
are systems with larger than average errorbars. The J and 
H plots also have little dispersion, to the extent that this can be assessed
from their smaller number of measurements. The V band scatter 
is  much larger than the measurement errors, which on average are 
actually somewhat smaller for V than for K. This different behaviour
of the visual and infrared bands, first seen so clearly here, 
is predicted by all theoretical models,
as discussed for instance in the recent review by Chabrier \& Baraffe 
(\cite{chabrier00}). It results from the metallicity dispersion of the 
solar neighbourhood 
populations, through the interplay of two physical mechanisms:
\begin{itemize}
\item a larger metallicity increases the atmospheric opacity in the visible
range, which is dominated by TiO and VO molecular bands. For a given 
bolometric luminosity it therefore shifts the flux distribution 
towards the infrared;
\item a larger metallicity decreases the bolometric luminosity for a given
mass.
\end{itemize}
In the visible bands both effects work together,
to decrease the visible luminosity of the more metal-rich stars
at a given mass.
In the near-infrared on the other hand, the redward shift of the flux
distribution of the metal-rich stars counteracts their lower bolometric
luminosity. The models therefore predict that infrared absolute magnitudes
are largely insensitive to metallicity, and our empirical M/L relations 
confirm this. 

At visible wavelengths on the other hand, metallicity determinations 
now become a crucial limiting factor in accurate comparisons with 
stellar models, as has long been
the case for more massive stars (e.g. Andersen \cite{andersen91}). 
Quantitative metallicity measurements of M dwarfs are unfortunately 
difficult in the optical range (e.g. Valenti et al. \cite{valenti98}), but 
near-IR spectroscopy 
offers better prospects (Allard, private communication).
GJ~2069A and Gl~791.2 represent spectacular illustrations of the intrinsic
dispersion of the V band M/L relation, as already discussed in their 
respective discovery paper (Delfosse et al. \cite{delfosse99a}; 
Benedict et al. \cite{benedict00}): 
these four stars are underluminous
by $\sim$2~magnitudes for their masses, compared to solar metallicity models 
and to other stars. We recently discovered an additional faint component
(Beuzit et al., in prep.) in GJ~2069A, which, if anything, 
further slightly increases its
distance from a solar metallicity M/L. This discrepancy is best explained
if the Gl~791.2 and GJ2069 systems are metal-rich by $\sim$0.5~dex. 
Their near-IR
absolute magnitudes should be much more consistent with the average
relations, but have not yet been measured. 
%

\section{Comparison to theoretical models}

\begin{figure*}
\begin{tabular}{cc}
\tabcolsep 0.2mm
\psfig{height=6.2cm,file=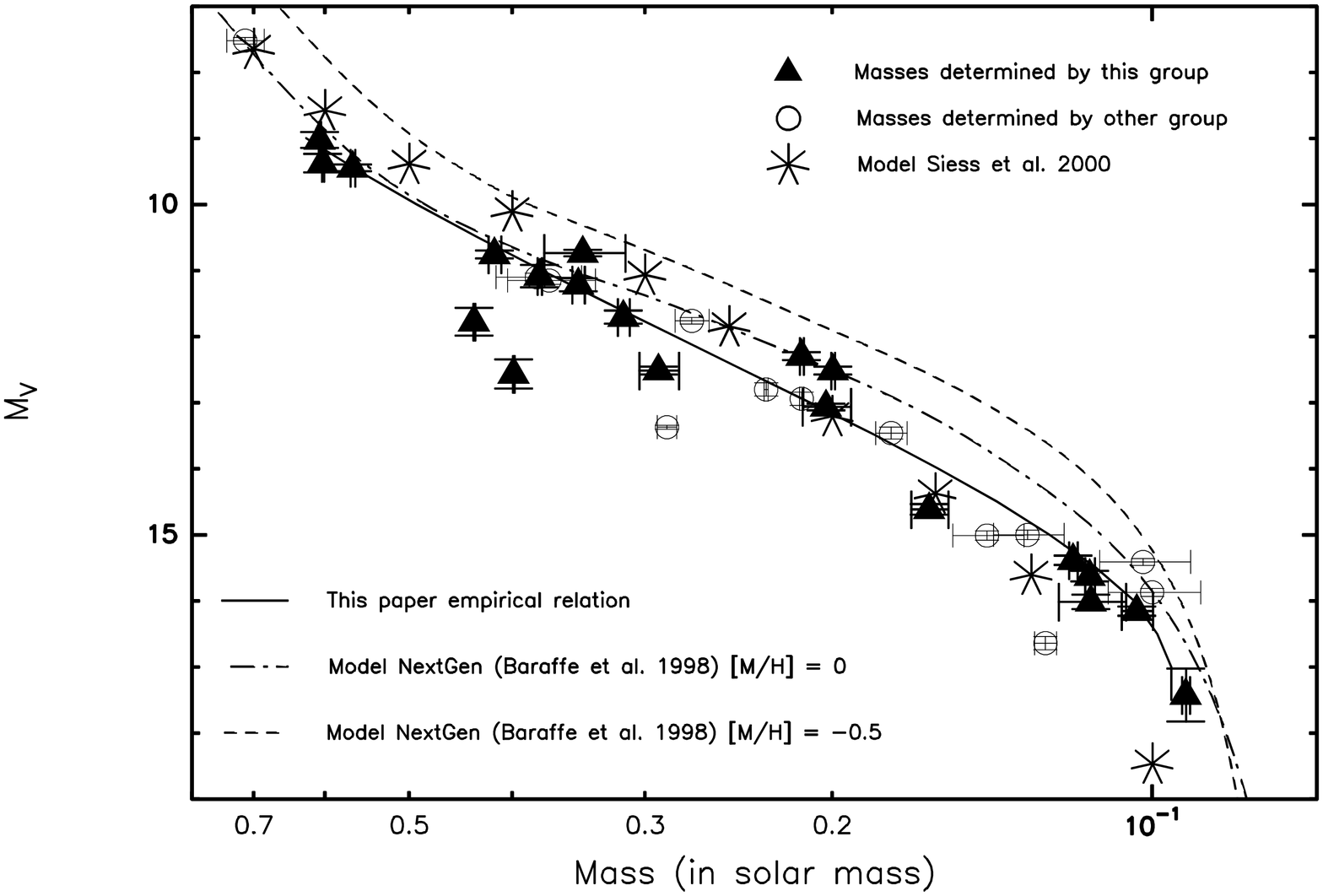,angle=-90} &
\psfig{height=6.2cm,file=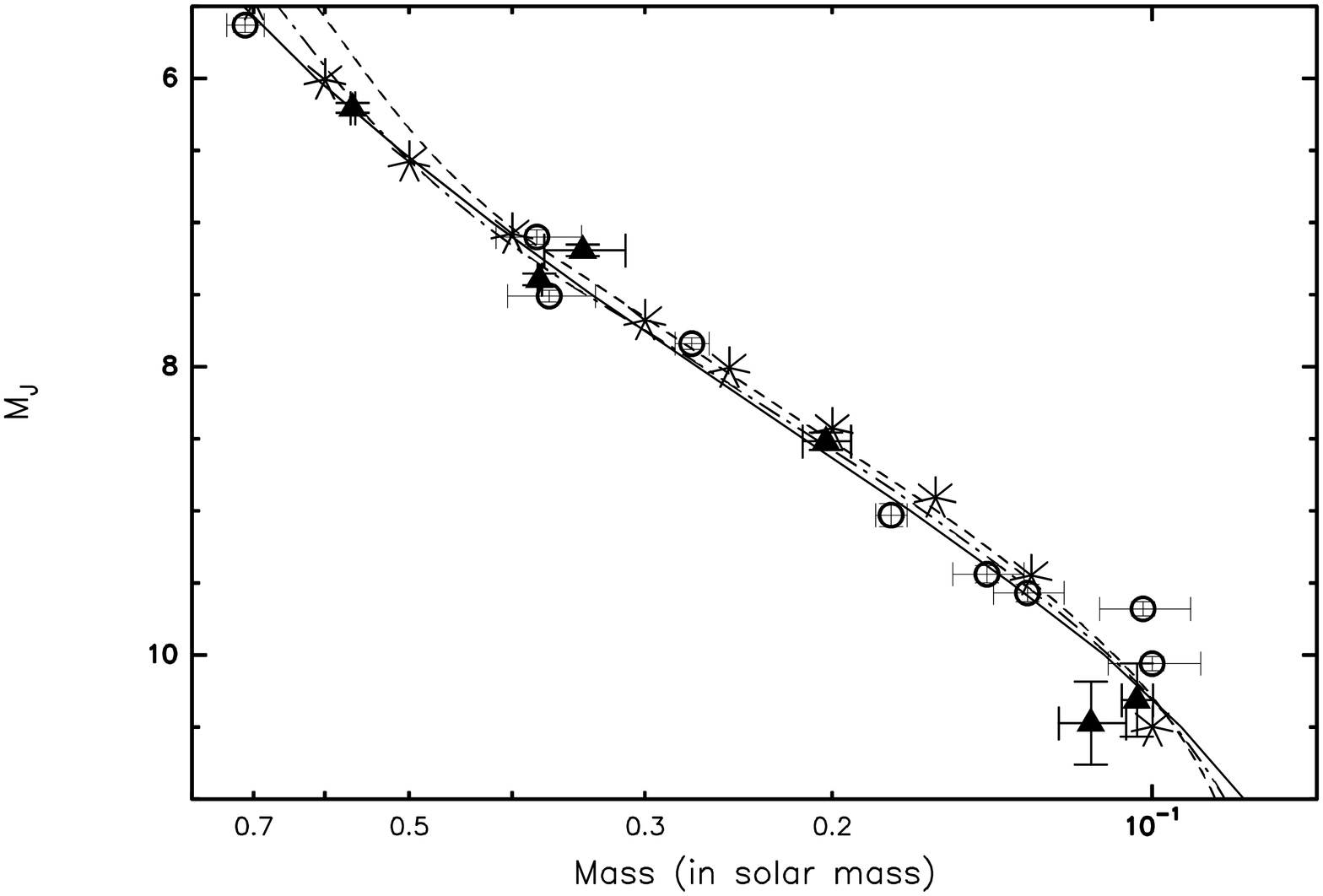,angle=-90} \\
\psfig{height=6.2cm,file=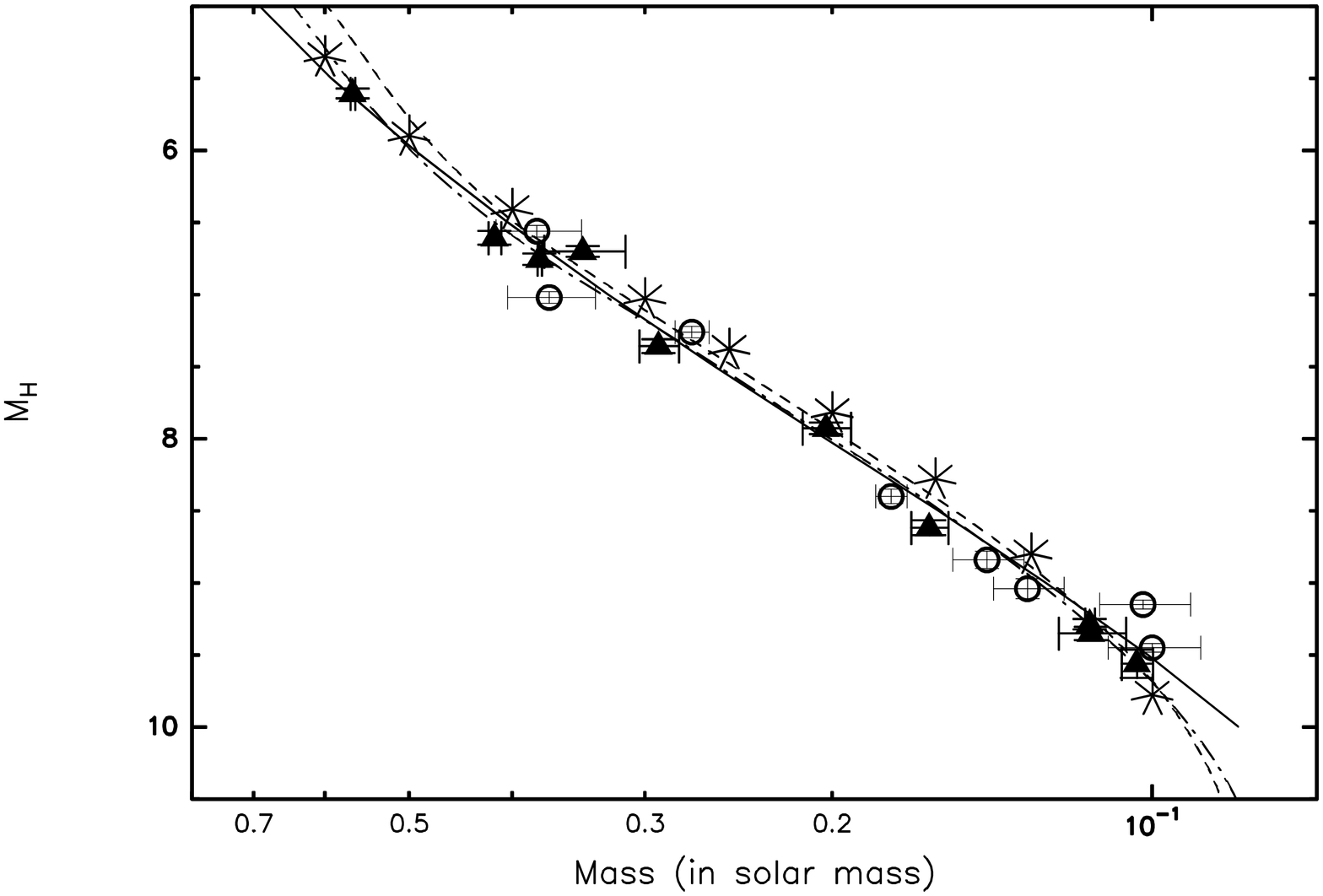,angle=-90} &
\psfig{height=6.2cm,file=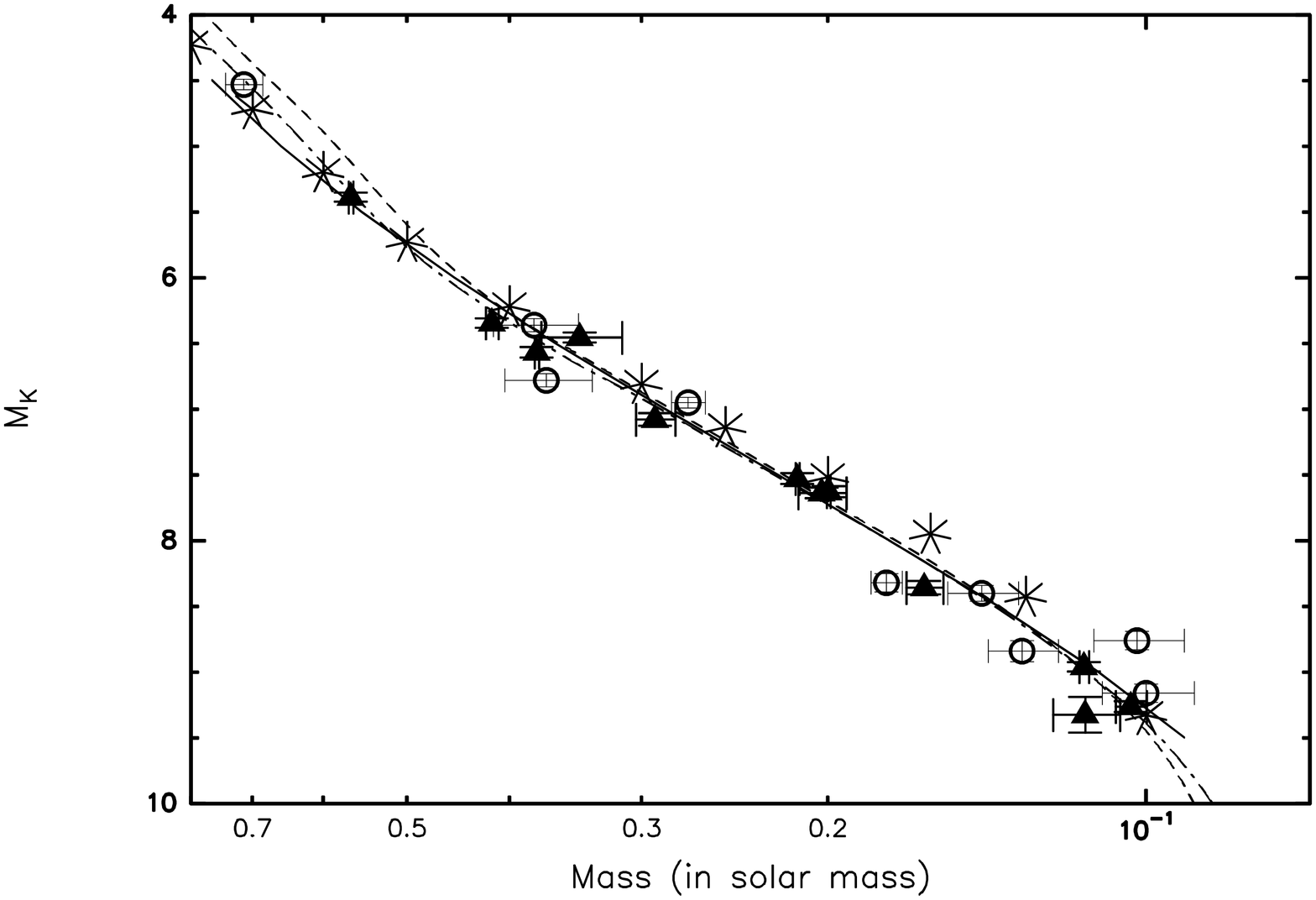,angle=-90} \\
\end{tabular}
\caption{comparison between V, J, H and K band M/L 
observational relation and theoritical ones. 
The three curves are 5~Gyr theoretical isochrones from 
Baraffe et al. (\cite{baraffe98})
for two metallicities and our polynomial fit. The asterisks represent
5~Gyr solar metallicity models from Siess et
al. (\cite{siess00}).} 
\label{mass-theo} 
\end{figure*}

Figure~\ref{mass-theo} compares the empirical M/L data with the
corresponding 5 Gyr theoretical isochrones (for which even the lowest-mass
stars on the plot have settled on the main sequence) of Baraffe et al. 
(\cite{baraffe98}; 
hereafter BCAH) and Siess et al. (\cite{siess00}; hereafter 
SDF). For very low mass stars these two sets of models are up to now the 
only ones to use realistic model atmospheres as outer boundary conditions
to the stellar interior equations. This has been found necessary for
accurate results in this mass range (Chabrier et al. 
\cite{chabrier96}). BCAH use atmospheres from Hauschildt et al. 
\cite{hauschildt99}, while SDF use the older Plez (\cite{plez92}) models.
Besides this, and the different input physics that they use, the two sets
of models differ by the technique used to compute observational quantities.
SDF use the empirical bolometric correction tables of Kenyon \& Hartmann 
(\cite{kenyon95}) to deduce absolute magnitudes in the various photometric
bands from their theoretical bolometric luminosities and 
effective temperatures.
BCAH on the other hand adopt a purely {\em ab initio} approach, and
compute the absolute magnitudes from the stellar radii, the model atmosphere 
spectra, and the transmission profile of the photometric filters.

At the scale of Fig.~\ref{mass-theo} the BCAH and SDF models are nearly 
indistinguishable for the infrared bands, and both produce an impressive 
agreement there with the observational data.
This is particulary striking for the K band, where many measurements
define the M/L relation. The same is apparently true for J and H, where
more data would nonetheless be welcome to confirm this behaviour.

In the V band (Fig.~\ref{mass-theo}) on the other hand, neither of the
two sets of models reproduces the observations perfectly. The BCAH models and 
the observations agree well above $\sim$0.5~\Msol, though with significant 
dispersion, but they diverge somewhat for lower masses. Below 
0.2~\Msol, the solar metallicity models are systematically too luminous 
by $\sim$0.5 magnitude. As most of the bolometric flux of such stars 
emerges in the near-IR bands, where the agreement is excellent, the 
models necessarily provide a good account of the relation between mass 
and bolometric magnitude.
Their overall description of the stars is therefore most likely correct, 
and the V band discrepancy probably points to a relatively localized 
problem in the models. The two leading explanations for this
discrepancy (Baraffe \& Chabrier, private communication) are either
a V band opacity source that would be missing in the atmospheric
models, or some low level problem in the physical description of the
shallower atmospheric levels which emit the visible flux.

The SDF models by contrast are $\sim$0.5~mag too luminous in the V band 
above 0.3\Msol, produce an excellent agreement with the observations for 
0.2\Msol, and look sub-luminous for 0.1\Msol. Here again, the excellent 
near-IR agreement with the observations indicates that these models
nicely reproduce the relation between mass and bolometric magnitude,
and the discrepancy rests in the V band bolometric correction.
Indeed, the SDF models mostly target PMS stars and the Kenyon \& Hartmann 
(\cite{kenyon95}) bolometric correction that they use applies for 
T Tauri stars, which have lower gravity than main sequence 
stars and hence somewhat different colours. The use of observational 
bolometric correction, which by definition are unaffected by missing
opacity sources, on the other hand most likely explains why the SDF 
models agree better with the V band observations at $\sim$~0.2{\Msol}.

The characteristics of very low mass stars are frequently derived from
photometry in the redder CCD bands, R and especially I, where these
objects are brighter than in the V band. The validity of the theoretical
M/L relation for these red bands is thus of significant interest, but too 
few VLMSs with accurate masses have known luminosities in the R, I, or z 
filters to provide a fully empirical verification. One can 
note however that Baraffe et al. (\cite{baraffe98}) observe that below 
T$_{\rm eff}~\sim$~3700~K the BCAH model $V-I$ and $V-R$ colours are too 
blue by 0.5~mag at a given luminosity, while colours
which don't involve the V band are much better predicted. Since 
the BCAH model M$_{\rm V}$ are also $\sim$0.5 mag too luminous for their
mass, this suggests that the atmospheric models have a problem that
is specific to the V band. The model M/L relations for 
the R, I and z bands are then probably more nearly correct. If valid this 
inference would suggest that the root of the problem rests in the V
band opacity rather than in the physical description of the visible
photosphere, which would probably affect a broader wavelength range.

\section{Conclusions}

%

Empirical masses of 0.2 to 10\% accuracy validate the near-IR Mass/Luminosity
relations predicted  by the recent stellar models of Baraffe et al. 
(\cite{baraffe98}) and Siess et al. (\cite{siess00}),
down to $\sim$0.1~\Msol. They 
also point out some low level ($\sim$0.5~mag) deficiencies of these models 
in the V band.
Perhaps more importantly however, the V band M/L diagram represents 
direct evidence for an intrinsic dispersion around 
the mean M/L relation. This had previously remained hidden in the 
measurement noise, but there is, as theoreticians have kept telling us, 
no such thing as one
single M/L relation for all M dwarfs. This is particularly true for 
the visible bands, while the dispersion in the near-IR JHK bands
is much lower. Comparisons between measured masses and theoretical
models will therefore increasingly depend on metallicity measurements
for individual systems, which are not easily obtained.


The $\sim$0.5~magnitude discrepancy between observational and model 
masses derived from visible photometry has some consequences 
for mass functions determination. As mass cannot be determined for 
volume-limited samples, the mass function is always obtained from
a luminosity function, by writing that
\begin{displaymath}
{\frac{{\rm d}N}{{\rm d}M}}~={\frac{{\rm d}N}{{\rm d}L}}{\times}{\frac{{\rm d}L}{{\rm d}M}}
\end{displaymath}
and the slope of the M/L relation therefore plays a central role in
its derivation. Below 0.5{\Msol} the d$L/$d$M$ slope of the empirical 
M/L relation is steeper than that of the BCAH models and shallower than 
for the SDF ones, by 10 to 20\%.  Their use will therefore respectively 
underestimate and overestimate the number of lower mass stars by 
this amount.
Probably more seriously, the large dispersion around the V band 
M/L relation will introduce large Malmquist-like biases in the derived mass 
function, which would need an excellent characterization of this
dispersion to be corrected. The infrared relations have both better 
agreement with the observations and much lower dispersion. We strongly 
recommend that they be used rather than the V band relations, whenever 
possible.


\begin{acknowledgements}

We thank the technical staffs and telescope operators of both OHP and CFHT
for their support during the long-term observations which have led to these
results. We are grateful to Gilles Chabrier, Isabelle Baraffe, 
France Allard and Maria Rosa Zapatero Osorio for many useful discussions.
We are also indebted to Lionel Siess for providing his models
prior to publication.

\end{acknowledgements}

\end{document}